\documentclass[english,reprint,amsmath,amssymb,prd,showpacs]{revtex4-1}
\usepackage[latin9]{inputenc}
\usepackage{amsmath}
\usepackage{graphicx}

\makeatletter

\DeclareFontEncoding{LGR}{}{}
\DeclareRobustCommand{\greektext}{%
  \fontencoding{LGR}\selectfont\def\encodingdefault{LGR}}
\DeclareRobustCommand{\textgreek}[1]{\leavevmode{\greektext #1}}
\ProvideTextCommand{\~}{LGR}[1]{\char126#1}

\usepackage{dcolumn}
\usepackage{bm}

\makeatother

\begin{document}

\title{Machian derivation of the Friedmann equation}

\author{{\normalsize{}Herman Telkamp}}

\address{Jan van Beverwijckstraat 104, 5017 JA Tilburg, The Netherlands}
\email{herman\_telkamp@hotmail.com}


\date{June 9, 2016}
\begin{abstract}
\noindent Despite all fundamental objections against Newtonian concepts
in cosmology, the Friedmann equation derives from these in an astoundingly
simple way through application of the shell theorem and conservation
of Newtonian energy in an infinite universe. However, Friedmann universes
in general posses a finite gravitational horizon, as a result of which
the application of the shell theorem fails and the Newtonian derivation
collapses. We show that in the presence of a gravitational horizon
the Friedmann equation can be derived from a Machian definition of
kinetic energy, without invoking the shell theorem. Whereas in the
Newtonian case total energy translates to curvature energy density,
in the Machian case total energy takes on different identities, depending
on the evolution of the horizon; we show that in the de Sitter universe
Machian total energy density is constant, i.e. appears as cosmological
constant.
\end{abstract}

\maketitle

\section{Introduction: \protect \\
Newtonian derivation of the Friedmann equation}

The Friedmann equation, describing the evolution of the scale factor
$a$ of the expanding universe, reads

\begin{equation}
\frac{\dot{a}^{2}}{a^{2}}=\frac{8}{3}\pi G(\rho+\rho_{k}),\label{eq:Fr}
\end{equation}
where, according general relativity (GR), $\rho$ is the total density
of the various matter sources and where $\rho_{k}\!\propto\!a^{-2}$
represents curvature energy. 

The density of pressureless matter (dust) causes deceleration of expansion,
which is commonly attributed to the attractive force of gravity between
cosmic masses. This view on pressureless matter, i.e., deceleration
by attraction, is supported by the well known Newtonian cosmology
(see e.g. \citep{Liddle,Rindler}), which involves both the shell
theorem \textit{(the field within an empty spherical mass shell is
uniformly zero)} and the conservation of energy, $E=T+V$, of a test
mass $m$ at the edge of an arbitrary spherical volume of the universe,
where $T$ and $V$ represent Newtonian kinetic and potential energy
of $m$ relative to (the center of) this sphere. For a spherical volume
of proper radius $R_{s}=a\chi_{s}$, where $\chi_{s}$ is the constant
comoving coordinate, this straightforwardly leads \citep{Liddle}
to the Newtonian energy conservation equation 
\begin{equation}
T+V\,=\,{\textstyle \frac{1}{2}}m\dot{a}^{2}\chi_{s}^{2}\,\,-{\textstyle \frac{4}{3}}m\pi G\rho a^{2}\chi_{s}^{2}\,\,=E=const.,\label{eq:Newton energy}
\end{equation}
which is precisely the Friedmann equation (\ref{eq:Fr}); $\chi_{s}$
is fixed, so constant total energy $E$ indeed translates to curvature
density, $\rho_{k}\!\propto\!E/a^{2}\chi_{s}^{2}$, as can be verified
easily. In a strict Newtonian sense the matter density $\rho$ regards
ordinary pressureless matter only, but for the Newtonian line of thought
there is no objection to also include other matter sources, like this
is essential in GR.

We note that Newton was surprisingly successful as the same equation
was derived by Friedmann from GR (which derivation is actually much
more involved, yet physically sound). What is bothering though is
that, considering all fundamental objections against Newtonian cosmology,
the Newtonian result is almost too good to be true. Not only the result
is consistent with the GR derivation, the Newtonian derivation also
shows that, with the simplest of arguments, classical concepts of
kinetic, potential and conserved total energy still seem to make sense,
while these are often believed not to hold globally in (relativistic)
cosmology. Perhaps the Newtonian idea of conservation of energy in
the universe - seemingly at war with GR - deserves more credit than
being educational alone. 

On the other hand, Newtonian concepts like infinite speed of gravity
and absolute space are generally considered unphysical. We therefore
assume gravity propagating at the speed of light, along with the presence
of a gravitational cosmic horizon, which is the case with general
Friedmann universes. However, as we will note, the application of
the shell theorem fails in the presence of a horizon. And so does
the Newtonian derivation. Instead, we consider purely relational (Machian)
definitions of energy, due to Schr\"{o}dinger \citep{Schroedinger,*[{ For an English translation see: }]Barbour}.
Applying these, we arrive at a Machian derivation of the Friedmann
equation by assuming conservation of total Machian energy, and without
invoking the shell theorem. This derivation holds, like in the GR
case, for general Friedmann universes, with or without a horizon. 

\section{Limitations and inconsistency of the Newtonian derivation }

Though intuitive (matter attracts other matter), Newtonian cosmology
rests on the dubious assumptions of instantaneous ``action at a distance''
in an infinite unbounded universe. Within GR however, Friedmann universes
in general posses a gravitational horizon, due to propagation of gravity
at the speed of light in a universe of finite age. We therefore assume
a propagating gravitational horizon at comoving distance $\chi_{g}$.

\begin{figure}[h]
\hspace*{-1.2cm}\includegraphics[bb=2cm 8.5cm 17cm 20.5cm,clip,scale=0.4]{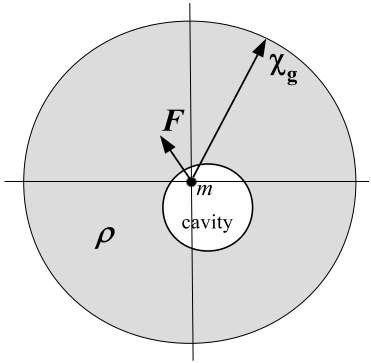}\caption{{\small{}Gravitational force $F$ inside an empty spherical cavity
within the horizon}\textit{\small{} $\chi_{g}$.\label{fig:Gravitational-force-}}}
\end{figure}

The whole argument of Newton's derivation is based on the assumption
that the mass distribution outside an empty spherical cavity of the
universe is spherically symmetric, so that the shell theorem applies
and the field inside the cavity is uniformly zero. This assumption
holds in an infinite homogeneous isotropic universe. However, the
presence of a gravitational horizon, centered at an observer somewhere
in the cavity, breaks this symmetry; due to the introduction of a
horizon the field inside the cavity is no longer uniformly zero, as
is clear from Fig.\ref{fig:Gravitational-force-}; the asymmetry causes
a non-zero field directed away from the center of the cavity.

As a result the Newtonian derivation collapses for universes bounded
by a gravitational horizon, which, ironically, is the case with general
Friedmann universes. And actually, one would not expect differently,
since due to symmetry the gravitational field in the perfectly homogeneous
isotropic universe is zero everywhere, \textit{with or without a horizon}.
This makes the concept of deceleration by gravitational attraction
questionable. For the same reason one must question the concept of
a ``repulsive force'' due to vacuum energy. 

With no forces acting, what then is left to even explain deceleration
in a Newtonian sense? The zero field points at constancy of the kinetic
energy of recession; no energy is being exchanged. In a Newtonian
context, where mass inertia is invariant, constant kinetic energy
implies constant recession velocities, i.e., a coasting universe,
as in Melia \citep{Melia}. Hence, if Newtonian physics predicts anything,
then it is a coasting universe, rather than a decelerating universe;
a conclusion earlier arrived at e.g. by Layzer \citep{layzer}. 

\section{Machian physics}

The Machian approach due to Schr\"{o}dinger \citep{Schroedinger}
shares features of Newtonian mechanics, however defined in a purely
relational framework, by which it considers causal relations of all
matter within the cosmological horizon. Such an explicit treatment,
as pursued hereafter, discerns the Machian approach from both Newton's
and Einstein's theory. In a Machian context mass inertia is a relational
property which depends on the ``distribution of matter'', and so
inertia is likely to evolve in an expanding universe with a cosmological
horizon. Like in GR, and following Schr\"{o}dinger approach in his
Machian derivation of the anomalous perihelion precession \citep{Schroedinger},
we assume the influence of the matter sources is through the gravitational
potential. That is, we assume the inertia of a point particle to depend
on the gravitational potentials of all other interacting point particles,
in a one-on-one fashion. By integration of the elementary point particle
relations one can derive expressions for the interaction of arbitrary
finite bodies, like e.g. two solid spheres, or (as in our case) a
small test particle relative to the surrounding sphere of cosmic matter,
as explained below.

How to actually determine (cosmic) gravitational energy is a matter
still open to debate. But given the surprising success of Newtonian
cosmology, we assume a Newtonian potential. Though, we do allow the
density parameter $\rho$ to include the various matter sources. We
assume a homogeneous and isotropic universe. Different from Newton's
derivation, we consider test particle $m$ at the \textit{center}
of the cosmic sphere of comoving radius $\chi_{g}$, thus regard all
matter causally connected with $m$. The infinitesimal Newtonian potential
at the origin due to the cosmic matter element $\textrm{d}^{3}M$
at spherical coordinates $(\chi,\theta,\phi)$ is 
\begin{equation}
\textrm{d}^{3}\varphi(\chi,\theta,\phi)\,=\frac{-G\,\textrm{d}^{3}M}{a\chi}=-G\rho a^{2}\chi\sin\theta\,\textrm{d}\phi\textrm{d}\theta\textrm{d}\chi.\label{eq:d3Phi}
\end{equation}
Then the Newtonian cosmic potential due to all matter within the horizon
is 
\begin{equation}
\varphi\,=\int_{0}^{\chi_{g}}\!\!\!\int_{0}^{\pi}\!\!\int_{0}^{2\pi}\textrm{d}^{3}\varphi=-2\pi G\rho a^{2}\chi_{g}^{2}.\label{eq:Phi}
\end{equation}
This expression contrasts with Sciama \citep{Sciama}, who derived,
by the gravitoelectromagnetic analog of Maxwell's equations, a cosmic
potential equal to 
\begin{equation}
\varphi_{u}=-c^{2}.\label{eq:phi u-1}
\end{equation}
Remarkably, the potential $\varphi_{u}$ is independent of any cosmic
parameter. But as Sciama noted, it \textit{has} to be like that, given
that in our inertial frame Newton's laws hold without any reference
to the cosmic masses. It is tempting to assume both potentials to
be equal, $\varphi=\varphi_{u}$, but it is unnecessary for the present
derivation of the Friedmann equation to make such an assumption.

Note that the Newtonian gravitational potential is genuinely Machian
by its relational, frame independent nature. Newtonian kinetic energy,
on the other hand, is frame dependent, thus not relational, so clearly
not Machian. There is no established definition of Machian kinetic
energy. Therefore, before considering Machian kinetic energy of the
expanding universe, we will briefly introduce Machian physics according
to the approach taken by Schr\"{o}dinger in his Machian derivation
of the anomalous perihelion precession \citep{Schroedinger}: 

The main conceptual difference with Newtonian physics is that in Machian
physics according Schr\"{o}dinger inertia and kinetic energy are
not intrinsic properties of a particle, but are mutual properties
which arise from the interaction of the particle with all other (causally
connected) particles, just like this is true for potential energy
and the force of gravity. For the pair of point particles $m_{i}$
and $m_{j}$ we consider a frame independent definition of their mutual
Machian kinetic energy, according 
\begin{equation}
T_{ij}\equiv\tfrac{1}{2}\mu_{ij}\dot{r}_{ij}^{2}\,,\label{eq:Tij-1-1}
\end{equation}
where $r_{ij}$ denotes the proper radial distance (separation) of
the particles. Crucial here is that from a relational point of view
\textit{only} radial motion is meaningful in the physical relationship
of the two point particles. This ontological notion (epistemological
if you like) is due to Bishop Berkeley \citep{berkeley}, one of the
earliest critics of Newton, who first pointed out that any motion
between two point particles, other than their relative radial motion,
in otherwise empty space is \textit{unobservable}, therefore physically
meaningless, or inexistent for that matter. Thus, two particles in
circular orbit of each other have zero mutual kinetic energy between
them. Each particle does however have mutual kinetic energy relative
to the surrounding cosmic particles, namely proportional to the (square
of the) radial component of motion towards each cosmic particle. Thus
Machian physics is intimately connected with all the cosmic matter.

One can picture Machian kinetic energy between two particles as the
energy that would be dissipated if one would freeze the relative motion
between the particles. Fixing the separation of two spheres which
are in elliptic orbit would definitely affect kinetic energy, but
only as far as the radial component of motion is concerned (this also
explains why the anomalous perihelion precession depends on the amount
of eccentricity of the ellipse). To the contrary, fixing the distance
between two spheres in perfect circular orbit would not affect energies
at all; their mutual (Machian) kinetic energy is zero, just as Berkeley
argued. So, if there is zero kinetic energy between spheres in perfect
circular orbit, then the orbital kinetic energy of these two spheres
must be entirely due to the presence of cosmic matter. That is, due
to the motion of each sphere in or from the direction of surrounding
cosmic matter.

Like kinetic energy $T_{ij}$, Machian inertia $\mu_{ij}$ is a relational
and mutual property between any pair of point particles and is defined
as 
\begin{equation}
\mu_{ij}\,\equiv\,m_{i}\frac{\varphi_{j}(r_{ij})}{\varphi_{e\!f\!f}}\,=\,m_{j}\frac{\varphi_{i}(r_{ij})}{\varphi_{e\!f\!f}}\,=\,\frac{-Gm_{i}m_{j}}{\varphi_{e\!f\!f}\,r_{ij}},\label{eq:muij}
\end{equation}
where the effective potential 
\begin{equation}
\varphi_{e\!f\!f}={\textstyle \frac{1}{3}}\varphi\label{eq:Phi eff}
\end{equation}
serves as a normalization parameter which preserves consistency with
Newtonian inertia, as shown below. The reason that a factor $\frac{1}{3}$
of the cosmic potential $\varphi$ appears (see Schr\"{o}dinger \citep{Schroedinger})
is that in a Machian sense only the radial component of motion counts,
i.e., the two perpendicular components of motion do not contribute.
Hence, in any peculiar motion effectively only one third of the total
cosmic potential contributes to the kinetic energy between a mass
$m$ and the universe (see appendix \ref{App A} for details). 

Of course one can argue whether the above Machian definitions are
correct. Obviously they are no subtitute of GR. Yet, the underlying
physical concepts (observability, causal connection of all matter
within the cosmic horizon, abandoning absolute space, mutuality of
both inertia and kinetic energy) are ontologically as good as irrefutable.
But also the precise form of the definitions Eqs. (\ref{eq:Tij-1-1},\ref{eq:muij})
appears sensible. Inertia between two point particles is defined as
their mutual potential energy and so expresses the equivalence of
mass and energy. From these same definitions Schr\"{o}dinger straightforwardly
reproduced the GR expression of the anomalous perihelion precession.
Since the Machian approach explicitly considers the causal connection
of all matter within the gravitational cosmic horizon, it may provide
a useful alternative in the physical interpretation of cosmology,
like in the derivation of the Friedmann equation hereafter. 

\section{Machian derivation of the Friedmann equation}

Using the above elementary definitions {[}Eqs. (\ref{eq:Tij-1-1},\ref{eq:muij},\ref{eq:Phi eff}){]},
one can formulate the kinetic energy between a small mass $m$ at
the origin and all receding matter within the gravitational horizon
$\chi_{g}$. We assume (only initially) particle $m$ has a peculiar
velocity $v$. This to evaluate consistency of the Machian definitions
with Newtonian physics. For simplicity and without loss of generality
we assume this peculiar velocity is in the polar direction ($\theta=0$).
The radial velocity $\dot{r}$ between $m$ and cosmic matter element
$\textrm{d}^{3}M$ at coordinates $(\chi,\theta,\phi)$ is thus 
\begin{equation}
\dot{r}=\chi\dot{a}-v\cos\theta.\label{eq:drmdM}
\end{equation}
According Eq. (\ref{eq:muij}) the mutual inertia between $m$ and
matter element $\textrm{d}^{3}M$ is, using Eqs. (\ref{eq:d3Phi},\ref{eq:Phi}),
\begin{equation}
\textrm{d}^{3}\mu\,=\,m\,\frac{\textrm{d}^{3}\varphi(\chi,\theta,\phi)}{\tfrac{1}{3}\varphi(\chi_{g},a)}=m\,\frac{3\chi\sin\theta}{2\pi\chi_{g}^{2}}\,\textrm{d}\phi\textrm{d}\theta\textrm{d}\chi.\label{eq:d3mu}
\end{equation}
By definition Eq. (\ref{eq:Tij-1-1}) the kinetic energy between particle
$m$ and the cosmic matter element $\textrm{d}^{3}M$ is

\begin{equation}
\begin{array}{rl}
\textrm{d}^{3}T(\chi,\theta,\phi) & =\tfrac{1}{2}\textrm{d}^{3}\!\mu\,\dot{r}^{2}\,\\
\\
 & =m\,\frac{{\textstyle 3\chi\sin\theta}}{{\textstyle 4\pi\chi_{g}^{2}}}(\dot{a}\chi-v\cos\theta)^{2}\,\textrm{d}\phi\textrm{d}\theta\textrm{d}\chi.
\end{array}\label{eq:dTum-1-1}
\end{equation}

\noindent Then the total Machian kinetic energy between the particle
$m$ and all the cosmic matter within the horizon follows with some
math (see appendix \ref{App A}) from the integral

\begin{equation}
\begin{array}{rl}
T & =\tfrac{3}{4\pi}\frac{m}{\chi_{g}^{2}}{\displaystyle \int_{0}^{\chi_{g}}\!\!\!\!\int_{0}^{\pi}\!\!\!\int_{0}^{2\pi}\!\!\!\!\!}\chi\sin\theta(\dot{a}\chi-v\cos\theta)^{2}\:\textrm{d}\phi\textrm{d}\theta\textrm{d}\chi\\
\\
 & =\,\,{\textstyle \frac{3}{4}}m\dot{a}^{2}\chi_{g}^{2}\,+\,{\textstyle \frac{1}{2}}mv^{2}.
\end{array}\label{eq:Tm}
\end{equation}
Thus one obtains two distinct energies: recessional and peculiar kinetic
energy. The cross term of peculiar motion and recession vanishes due
to symmetry. We see that the peculiar part of the kinetic energy satisfies
the Newtonian definition and that the Newtonian inertia ($m$) is
being retrieved. This means that the equivalence principle is maintained
in peculiar motion, regardless of cosmic evolution. Hence, the Machian
principle remains unnoticed in the local frame. This is consistent
with the Hughes-Drever experiments \citep{Drever} (a null result
of anisotropy of inertia) and in agreement with both Dicke \citep{Dicke}
and Sciama \citep{Sciama}. Indeed, Newtonian physics makes no reference
to cosmological parameters, \textit{even} though it derives here from
a cosmological context.

Confining ourselves again to recessional motion only, we consider
a unit test mass $m$ at rest in the Hubble flow ($v\!=\!0$), and
drop the symbol $m$ for simplicity. The potential energy of the particle
is $V\!=\varphi$ {[}Eq. (\ref{eq:Phi}){]}, thus together with recessional
kinetic energy $T=\tfrac{3}{4}\chi_{g}^{2}\dot{a}^{2}$ we have the
Machian energy equation 
\begin{equation}
\tfrac{3}{4}\chi_{g}^{2}\dot{a}^{2}\,-2\pi G\rho a^{2}\chi_{g}^{2}\,=\,E.\label{eq:Mach energy eq}
\end{equation}

\noindent Similar to the translation of total energy $E$ to curvature
energy density $\rho_{k}$ in the Newtonian case, we introduce the
(Machian) total energy density parameter
\begin{equation}
\rho_{E}=E/2\pi Ga^{2}\chi_{g}^{2},\label{eq:rhoE-2}
\end{equation}
which, interestingly, is not necessarily curvature energy density,
as $\chi_{g}$ in general evolves with the scale factor $a$ (see
next section). Eq. (\ref{eq:Mach energy eq}) can now be rewritten
\begin{equation}
\tfrac{3}{4}\chi_{g}^{2}\dot{a}^{2}\,=\,2\pi G(\rho+\rho_{E})a^{2}\chi_{g}^{2}.\label{eq:Ext energy}
\end{equation}
Eliminating the common factor $\chi_{g}^{2}$, we obtain the Friedmann
equation 
\begin{equation}
\frac{\dot{a}^{2}}{a^{2}}=\frac{8}{3}\pi G(\rho+\rho_{E}).\label{eq:Friedmann Machian}
\end{equation}

\section{The identity of Machian total energy density}

The Machian energy equation Eq. (\ref{eq:Mach energy eq}) is nearly
identical to the Newtonian version Eq. (\ref{eq:Newton energy}),
however with a notable difference: $\chi_{s}$ is an arbitrary \textit{fixed}
comoving radius, while $\chi_{g}$ is (in general) an evolving gravitational
horizon, $\chi_{g}=\chi_{g}(a)$. Thus in the Newtonian case total
energy $E$ appears as curvature energy density \textit{always}, i.e.,
$\rho_{k}\propto a^{-2}\chi_{s}^{-2}$, while in the Machian case
total energy density in general also evolves with the horizon, $\rho_{E}\propto a^{-2}\chi_{g}^{-2}(a)$.
Therefore, quite intriguingly, $\rho_{E}$ may take identities different
from curvature energy, depending on the particular evolution of the
horizon. For instance, in the de Sitter universe, and so in the late
phase of the \textgreek{L}CDM model, the proper distance to the event
horizon is constant, hence $\chi_{g}\propto a^{-1}$. This translates
total energy density to constant vacuum energy density, i.e., $\rho_{E}=\rho_{\Lambda}=const$.
Thus Machian cosmology potentially gives physical interpretation to
the origin of the cosmological constant. 

\section{Conclusion}

We argued that application of the shell theorem fails with universes
bounded by a gravitational horizon, like general Friedmann universes
are. Consequently the Newtonian derivation of the Friedmann equation
collapses in the presence of a horizon, while the GR derivation holds
for general Friedmann universes. We showed that, alternatively, the
Friedmann equation can be derived from Machian arguments: for arbitrary
homogeneous isotropic universes, either with or without a horizon,
the Friedmann equation follows from conservation of a Machian definition
of energy, without invoking the shell theorem. We further showed that,
depending on the evolution of the horizon, the Machian total energy
density takes different identities. In particular in a de Sitter universe,
Machian total energy density appears as constant vacuum energy density,
thus provides interpretation to the cosmological constant. 

\appendix

\section{\label{App A}}

Calculation details of the integral Eq. (\ref{eq:Tm}), i.e., of the
Machian kinetic energy between a test mass $m$ at the origin and
all receding matter within the cosmic horizon $\chi_{g}$, where $m$
is assumed to have a peculiar velocity $v$ in the polar direction
$\theta=0$:

\begin{equation}
\begin{aligned}T & =\tfrac{3}{4\pi}\frac{m}{\chi_{g}^{2}}\int_{0}^{\chi_{g}}\!\!\!\!\int_{0}^{\pi}\!\!\!\int_{0}^{2\pi}\!\!\!\!\!\chi\sin\theta(\dot{a}\chi-v\cos\theta)^{2}\:\textrm{d}\phi\textrm{d}\theta\textrm{d}\chi\\
 & =\tfrac{3}{2}\frac{m}{\chi_{g}^{2}}\int_{0}^{\chi_{g}}\!\!\!\!\int_{0}^{\pi}\!\!\!\dot{a}^{2}\chi^{3}\sin\theta-2\dot{a}\chi^{2}v\sin\theta\cos\theta\\
 & \quad+\,\chi v^{2}\sin\theta\cos^{2}\theta\:\textrm{d}\theta\textrm{d}\chi\\
 & =\tfrac{3}{2}\frac{m}{\chi_{g}^{2}}\int_{0}^{\chi_{g}}\!\!\!\left[-\cos\theta\dot{a}^{2}\chi^{3}+0-{\scriptstyle \frac{1}{3}}\chi v^{2}\cos^{3}\theta\right]_{0}^{\pi}\:\textrm{d}\chi\\
 & =\tfrac{3}{2}\frac{m}{\chi_{g}^{2}}\int_{0}^{\chi_{g}}\!\!\!2\dot{a}^{2}\chi^{3}+{\textstyle \frac{2}{3}}\chi v^{2}\:\textrm{d}\chi\\
 & =\tfrac{3}{2}\frac{m}{\chi_{g}^{2}}\left[{\textstyle \frac{1}{2}}\dot{a}^{2}\chi^{4}+{\textstyle \frac{1}{3}}\chi^{2}v^{2}\right]_{0}^{\chi_{g}}\\
 & ={\textstyle \frac{3}{4}}m\dot{a}^{2}\chi_{g}^{2}+{\textstyle \frac{1}{2}}mv^{2}.
\end{aligned}
\label{eq:T det}
\end{equation}

Since only the radial component of motion ($\dot{r}=\dot{a}\chi-v\cos\theta$)
counts, the potential due to the cosmic matter element at coordinates
($\chi,\theta,\phi$) contributes to the peculiar (i.e., Newtonian)
kinetic energy of the object $m$ only by a fraction $\cos^{2}\theta$.
On average, i.e., in the spherical integral Eq. (\ref{eq:T det}),
this fraction gives rise to a factor 
\begin{equation}
\frac{\int_{0}^{\pi}\sin\theta\cos^{2}\theta\:\textrm{d}\theta}{\int_{0}^{\pi}\sin\theta\:\textrm{d}\theta}=\frac{1}{3}.\label{eq:1/3}
\end{equation}
Thus the effective contribution of the cosmic potential to the peculiar
inertia is only $\varphi_{e\!f\!f}=\frac{1}{3}\varphi$ {[}Eq. (\ref{eq:Phi eff}){]}.

\bibliographystyle{apsrev4-1}
%

\end{document}